\documentclass[aps,12pt,final,notitlepage,oneside,onecolumn,nobibnotes,nofootinbib,superscriptaddress,noshowpacs,centertags]{revtex4}

\usepackage{amsmath,amssymb,epsfig,placeins,subfigure,wrapfig,indentfirst,makeidx}
\usepackage[cp1251]{inputenc}
\usepackage[english]{babel}
\usepackage{amsfonts,amssymb}
\usepackage{mathrsfs}
\usepackage{dsfont}

\begin{document}

\title{Special features in the asymptotic expression of the electromagnetic spectrum from dipole which free fall into a Schwarzschild black hole} 


\author{A.A. Shatskiy\footnote{shatskiy@asc.rssi.ru}}
\affiliation{The Astro Space Center, Lebedev Physical Institute of
RAS, 84/32 Profsoyuznaya st., Moscow, 117997, Russia}

\author{I.D. Novikov}
\affiliation{The Astro Space Center, Lebedev Physical Institute of
RAS, 84/32 Profsoyuznaya st., Moscow, 117997, Russia}
\affiliation{The Nielse Bohr International Academy, The Nielse
Bohr Institute, Blegdamsvej 17, DK-2100 Copenhagen, Denmark}

\author{L.N. Lipatova}
\affiliation{The Astro Space Center, Lebedev Physical Institute of
RAS, 84/32 Profsoyuznaya st., Moscow, 117997, Russia}

\date{\today}

\begin{abstract}
The characteristic features were found in the electromagnetic spectrum of radiation from free falling dipole, when it is fall radially into a Schwarzschild black hole. 
These features can be used as another method for the black hole mass determinating.  
Also, these features can be used for the determination some characteristics of the magnetosphere or the accretion disk around the black hole. 

\end{abstract}

\maketitle

\section{Introduction}
\label{s1}

As is well known (see~\cite{Thorn1998, Novikov1986}) either black hole in general relativity characterized by only three parameters: mass, angular momentum and electric charge. 
All other features of the matter which collapsing into a black hole should be fully radiated. 
This property of black holes called "no hair theorem", due to this property all electromagnetic multipole moments should disappear as the system of charges close to the black hole horizon\footnote{If the black hole rotates and electrically charged (Kerr-Newman solution) it has its own magnetic moment.}. 
Here we consider consequences of this property for a Schwarzschild black hole, ie for non-rotating and non-charged. 
Let us consider the radiation arising during the accelerated motion of the dipole, which freely falling along the radius in to the black hole. 
Reverse influence of the particle on the black hole we ignore. 

To find the components of the field which radiated by an essentially non-periodic motion of the charge we can use two ways:

  1. Extracting square root from the relativistic bremsstrahlung power of the radiation charge 
(see~\cite{Ross1971} or~\cite{Landau1988}, \S 76). 

  2. The field transformation from a locally inertial reference system (related to the charge) to the Schwarzschild reference system at the point of observation (at infinity). 

Both methods give the same result. 
In our previous studies~\cite{Shatskiy2013, Shatskiy2013-2} we have described in detail the second method. 

Both of these methods differ from suggested earlier: 
solution of general relativistic wave equation (Dalamberta type) for the field --- 
see~\cite{Zerilli1970, Davis1971, Teukolsky1972}. 

The obtained of the spectrum solution indicated that the radiation peaks at frequencies about ${c/r_g}$ 
(here $c$ --- speed of light and $r_g$ --- radius of the event horizon of a black hole). 
So, for example, for a black hole in the center of the Milky Way with the black hole mass ${M_{bh}\sim 10^{6}M_\odot}$, almost all of the electromagnetic energy emitted at wavelengths ${\lambda_m\sim r_g\approx 10^{6} [km]}$. 
Such waves technically impossible to register on the Earth.
Moreover, even for a black hole with the mass order of the mass of the sun radiation peaks at a wavelength of about one kilometer, which is very complicates the registration of such waves. 

So a necessity arose to determine the asymptotic behavior --- in the electromagnetic spectrum from this radiation at high frequencies.   
If this asymptotic will be exponentially decreasing, then experimentally detecting this radiation will be technically impossible.  
And if the asymptotic will be have the power spectrum, with the small (in absolute value) exponents, it will be possible to register such radiation (in the "tail"\, spectrum), and it's characteristic features (of the radiation) for define the parameters of the black hole and characteristics dipole falling. 

As will be shown in Section~\ref{s4}, the asymptotic behavior of the spectral energy density of the radiation would be exactly a power law, with an exponent ${k=-2}$.

\section{The features of the dipole radiation spectrum} 
\label{s2}

\begin{figure*}
\subfigure[\label{R-1-1} ${r_0=2r_g}$]{\includegraphics[width=0.95\textwidth,  height=0.20\textheight]{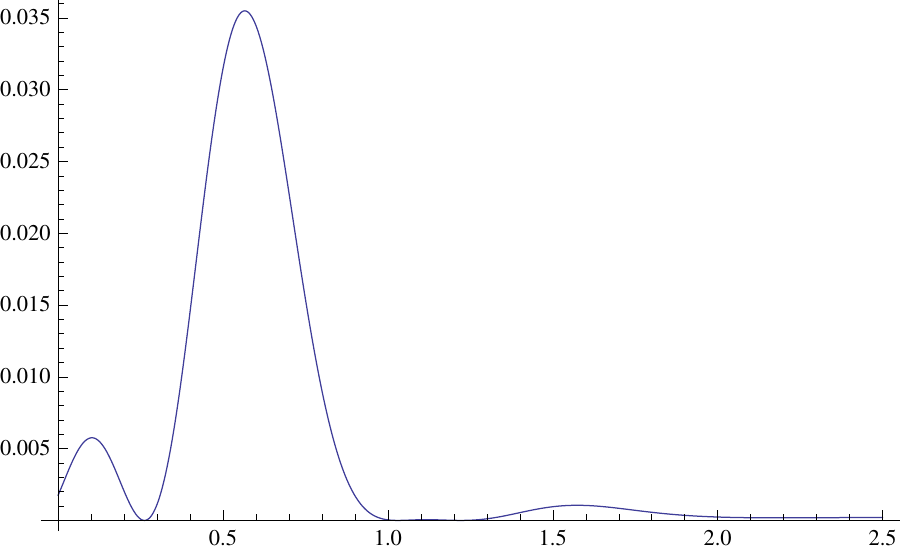}} 
\subfigure[\label{R-1-2} ${r_0=4r_g}$]{\includegraphics[width=0.95\textwidth, height=0.20\textheight]{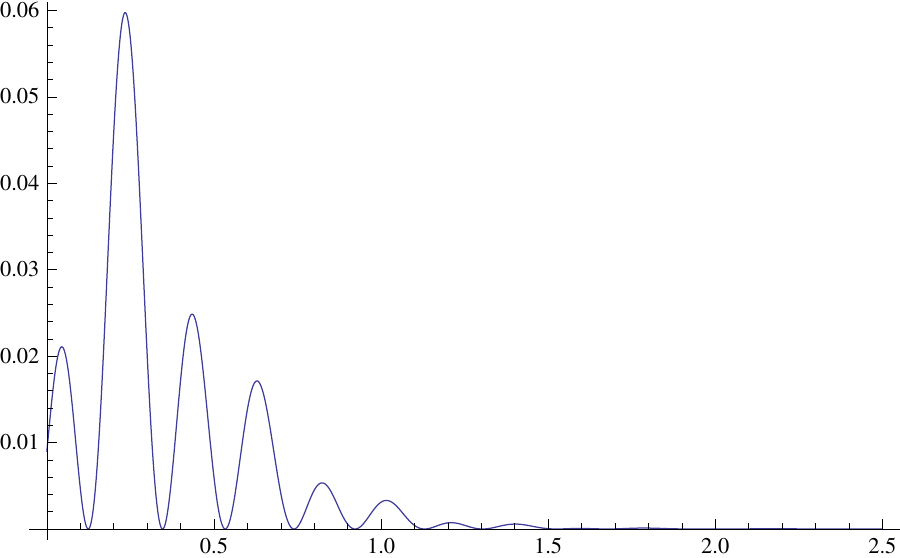}} 
\subfigure[\label{R-1-2} ${r_0=10r_g}$]{\includegraphics[width=0.95\textwidth, height=0.20\textheight]{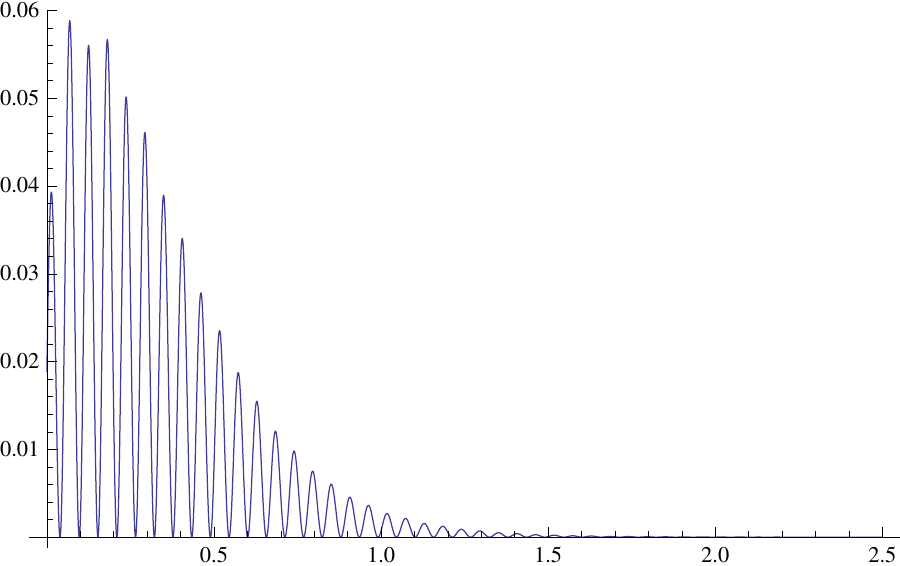}} 
\subfigure[\label{R-1-4} ${r_0=20r_g}$]{\includegraphics[width=0.95\textwidth, height=0.20\textheight]{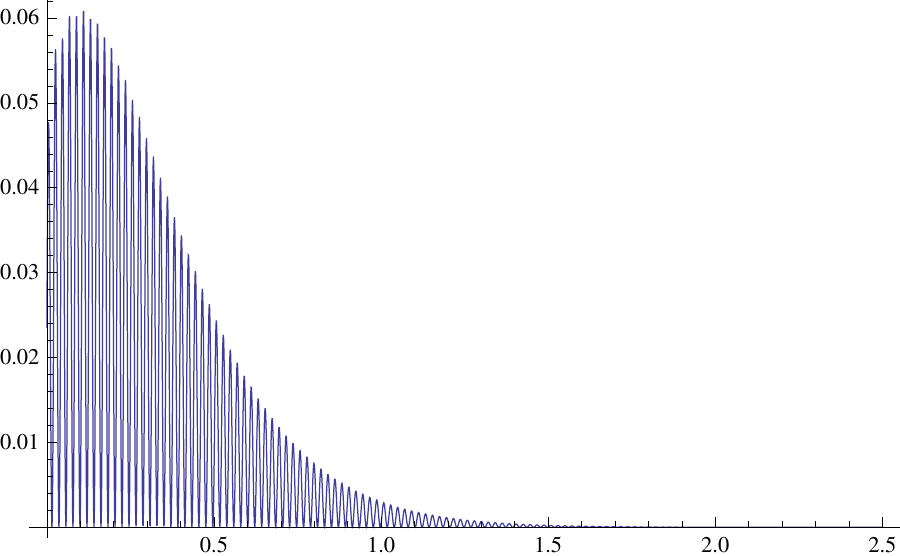}} 
\caption{{\label{R1}The dependence of the spectral density of the radiation energy ${\cal E}^d_{,{\omega}}$ --- in units of 
${d_0^2/(48\pi c r_g^2)}$, here argument ${({\omega}r_g/c)}$ with the different initial radii $r_0$ of falling dipole.}} 
\end{figure*}

The characteristic feature of the spectral energy density of the radiation ${\cal E}^d_{,{\omega}}$ for dipole falling into the black hole is its periodic nonmonotonic (damped oscillations --- see Fig.~\ref{R1}). 
And the corresponding frequency interval ${\Delta {\omega}}$ between the local extremums of the spectrum is the order
${c/r_0}$, where $r_0$ --- initial radius at which the dipole begins radial falling from rest. 
The cause of these oscillations of the spectrum is obvious non-periodicity at the velocity of the dipole (the particle moves only in one direction along the radius) and the fact that the movement begins from a finite radius --- see further.  
 
As is well known the spectral energy density is the square of the Fourier components of the field 
(see~\cite{Landau1988}, \S 66), so it is a positive definite function. 
At the same the Fourier components of the radiated field oscillate similarly at the same time decreasing by amplitude --- see Fig.~\ref{R1}. 

If the value ${r_0/r_g}$ tends to infinity (ie fall occurs from infinite radius), then interval between oscillations of frequency spectrum ${\Delta{\omega}}$ tends to zero, and the oscillation frequency tends to infinity. 

Therefore at infinity this spectrum is a monotonically decreasing function\footnote{Since the average value of a sine wave insquare is equal to the ${1/2}$.}, which asymptotic behavior will be found in the following section. 
In this case near the frequency ${{\omega} \sim c/r_g}$ the spectrum has the characteristic shape of a bell, which coincides with the results of~\cite{Zerilli1970, Davis1971, Teukolsky1972, Yakovlev1975, Dymnikova1977}. 
In these papers the spectra calculations were performed only for the case ${r_0/r_g \to  infty}$, and graphs of the spectra were developed only for the quadrupole radiation (and higher order multipoles).   
In works~\cite{Shatskiy2013, Shatskiy2013-2}, we have shown that the radiation of the dipole (which falling into a black hole) in the main approximation is dipolar. 
 
In this paper we consider the more general case ${r_0/r_g\sim 10}$ 
(not infinite --- as in the works~\cite{Zerilli1970, Davis1971, Teukolsky1972, Yakovlev1975, Dymnikova1977}). 
As already mentioned, in the general case, the spectral energy density has an infinite number of local maxima and local minima (equal to zero). 
This distinguishes it from the degenerate (asymptotic) case ${r_0/r_g\to\infty}$, has only one local maximum at the point  ${{\omega}\sim c/r_g}$. 

Due to the property of quasi-periodicity the spectrum for the general case ${r_0/r_g \sim 10}$ will be, in principle, to distinguish from the other spectra and analyze. 

\begin{figure*}
\subfigure[\label{Martel} at ${r_0/r_g = 40}$ --- from work~\cite{Martel2008}.
]{\includegraphics[width=0.49\textwidth, height=0.2\textheight]{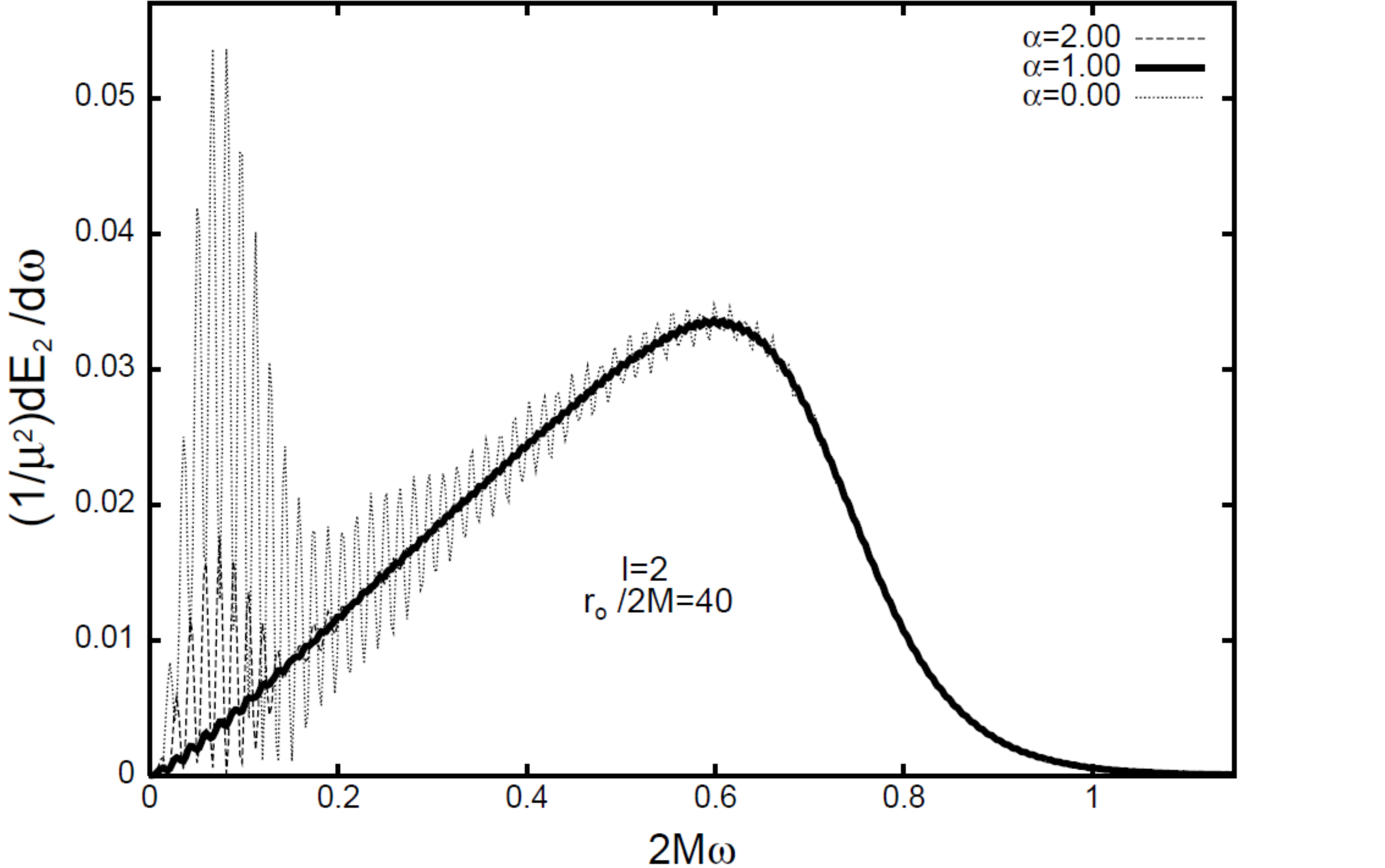}} 
\subfigure[\label{Davis} at ${r_0/r_g =\infty}$ --- from work~\cite{Davis1971}.
]{\includegraphics[width=0.49\textwidth, height=0.2\textheight]{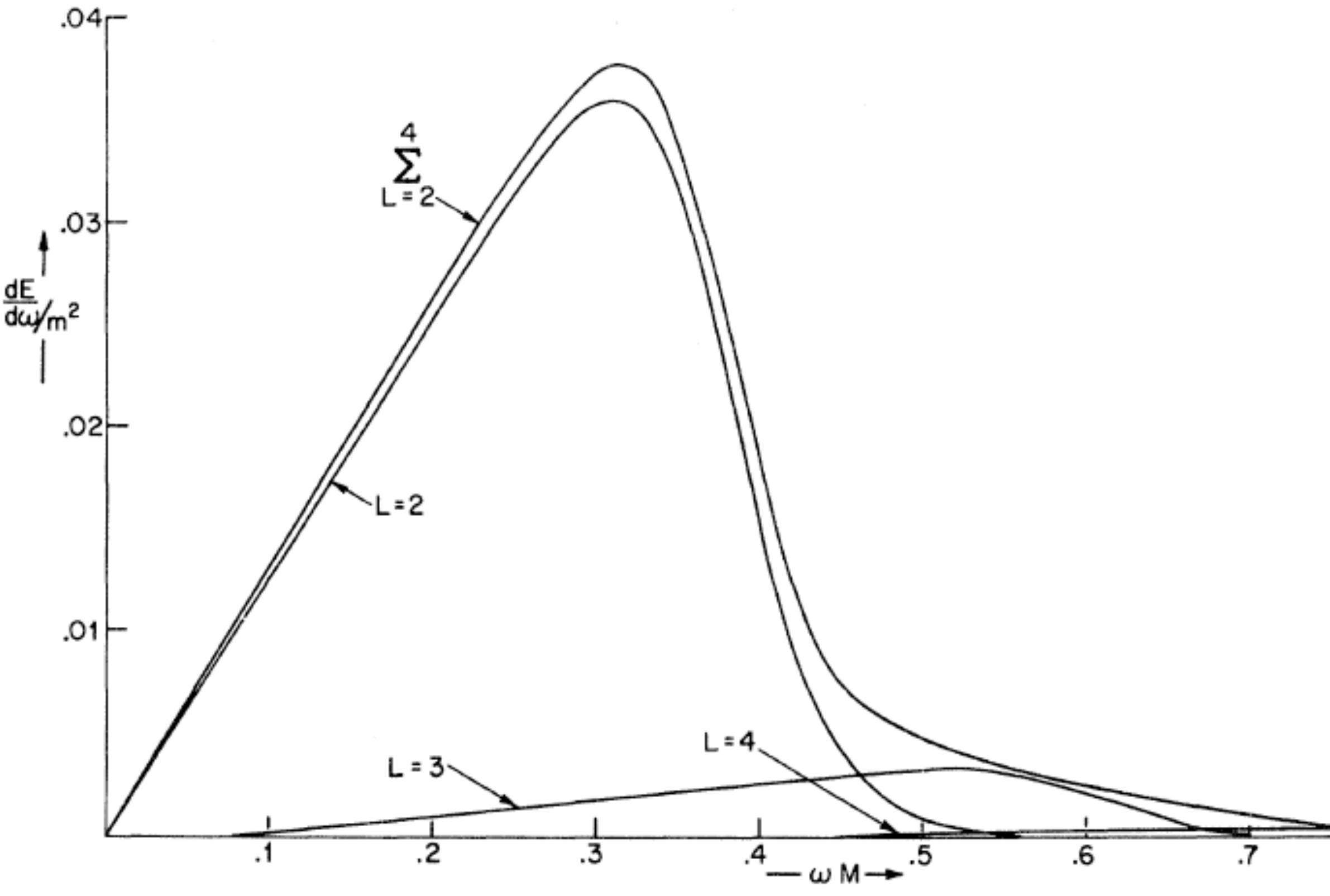}} 
\caption{{\label{R2}The dependence of the spectral density of radiation energy for the quadrupole gravitational radiation.}}
\end{figure*}

The results obtained for the dipole radiation, can be compared with the shape of the spectra from 
works~\cite{Davis1971, Martel2008}. In this works was obtained the spectr of quadrupole gravitational radiation for the test particle which radial falling into a Schwarzschild black hole --- see Fig.~\ref{R2}.   
Moreover in~\cite{Martel2008} (just like us) was considered the falling from a finite radius ${r_0=40r_g}$. 
And their spectrum contains the oscillations which similar to our --- see Fig.~\ref{Martel}. 

But the authors of~\cite{Martel2008} concluded that these oscillations can be caused by incorrect parameter selected in the there numerical method of calculation (calculation of these authors based on the Zerlini functions method).  
Regarding the work~\cite{Davis1971}: there is a falling from infinity and therefore there is no oscillation --- see Fig.~\ref{Davis}. 
But that is typical for the presented is also in this Fig.~\ref{Davis} spectra octupole ${(L = 3)}$ and hexadecapole 
${(L = 4)}$ radiation --- following multipole amplitude an in order smaller than the previous multipole. 
Hence we see that the amplitude of the dipole radiation should be much larger than the amplitude of all other multipoles (which we ignore in this paper).  

Where are the oscillations in the spectrum? 
To answer this question we point out that oscillation spectrum also appear when we calculating of the Fourier-transform from a non-periodic motion of a charge in a flat space (eg acceleration along a straight line for a finite time period).
In this case the interval between the frequency of the spectrum oscillations ${\Delta\omega = \pi /\Delta t}$ decreases to zero at infinite time interval velocity change (as well as in our problem). 
This simple model problem can easily be solved analytically (for the flat space-time). 
Thus the Fourier-transform from acceleration ${a(t)}$ has the form: 
${{\bf a_\omega} = a_0\left[\sin(\omega\Delta t) - \cos(\omega\Delta t) + 1 \right]/\omega}$. 
This is the cause of the oscillation spectrum.

\section{Asymptotics of the radiation spectrum at high frequencies} 
\label{s4} 

We consider the process of dipole radial falling (as a point, test particle) along a geodesic world line to the Schwarzschild black hole. 

In the works~\cite{Shatskiy2013, Shatskiy2013-2} we have found radiated tangential electric and magnetic fields which generated by charged particle (or dipole) in the black hole.

These fields have been found due to the relativistic field transformations from locally inertial reference frame connected with the moving charge (or dipole) to a fixed Schwarzschild reference frame of the observer. 
After this was calculated the Fourier transformatons of the found radiation fields. 
The radiation spectrum is Fourier transform in the square see~\cite{Landau1988}, \S 66. 

We introduce some notation:  
\begin{eqnarray} 
a\equiv\sqrt{1-r_g/r_d} ,\quad 
b\equiv\sqrt{r_g/r_d - r_g/r_0} ,\quad 
c_1\equiv\sqrt{1-r_g/r_0}\, .  
\label{abc-1} \end{eqnarray} 
Let us write down the expressions for the Fourier transform of the field from works~\cite{Shatskiy2013, Shatskiy2013-2}: 
\begin{eqnarray} 
{\bf F_{{\omega}d_\perp}^{\theta t}} = 
\int\limits_{0}^{\infty} \frac{F^{\theta t}_{d_\perp}\, \left[\sin ({\omega} t) + \cos ({\omega} t)\right]}{2}\, dt\, ,   
\label{4-2-1} \\
F^{\theta t}_{d_\perp} = \frac{d_0}{r^2}\cdot \frac{r_g a^2 (b+c_1) 
[a^2 c_1 - 2b^2 (b+c_1)]}{4 c_1 r_d^3 [(b+c_1)^2 - a^4 b^2]^{3/2}} \, .  
\label{4-2-2}
\end{eqnarray} 
Function ${F^{\theta t}_{d_\perp}{}_{(t,r)}}$ is a tangential electric field of the dipole radiation $d_0$ at the time of its taking away $r_d$ from the black hole (for the transverse orientation of the dipole).

And the dipole law of motion ${t(r_d)}$ defined by the expression which obtained in the same 
papers~\cite{Shatskiy2013, Shatskiy2013-2}: 
\begin{eqnarray}
c t(r_d) =  \sqrt{r_d(r_0-r_d)(r_0/r_g-1)}
+ \frac{(r_0+2r_g)\sqrt{r_0/r_g-1}}{2}\cdot \arccos\left( \frac{2r_d}{r_0} - 1\right)  + \nonumber \\
+ r_g\cdot\ln \left[ \frac{2\sqrt{r_d(r_0-r_d)(r_0/r_g-1)} +r_0 +r_d(r_0/r_g-2)}{r_0(r_d/r_g-1)} \right] 
\label{4-3}
\end{eqnarray} 
Thus, the initial conditions correspond to the ${t{(r_d=r_0)}=0}$. 

Let's split the integral (\ref{4-2-1}) into two parts: (from $0$ to $t_1$) and (from $t_1$ to infinity). 
Let the time $t_1$ corresponds to the position of the dipole at the radius $r_1$. 
Let us first consider the second integral (from $t_1$ to infinity).\\ 
Let us denote: ${\delta (r_d)\equiv (r_d/r_g -1)}$ and ${\delta_1\equiv (r_1/r_g - 1)}$. \\ 
Select the radius $r_1$, so that ${\delta \le \delta_1 <<1}$ (it corresponds to ${t_1>>r_g/c}$).

\begin{figure*}
\includegraphics[width=0.95\textwidth, height=0.3\textheight]{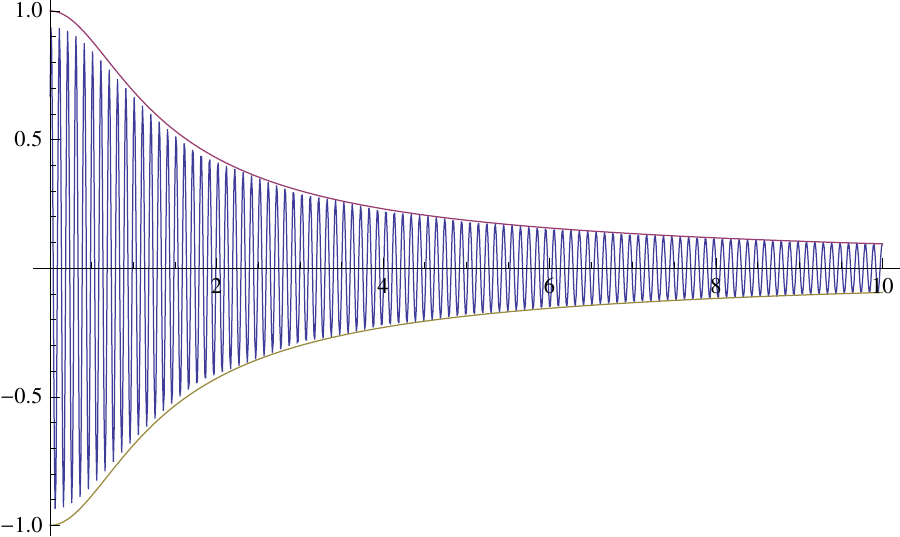}
\caption{{\label{R3}Graph for the second integral (\ref{4-2-1}) (from $t_1$ to infinity) --- part of the Fourier transform of the field $F^{\theta t}_{d_\perp}$ and graph of attractor (\ref{4-7}) at ${r_0=10r_g}$ and ${\delta_1=10^{-3}}$. 
The vertical axis is scaled in units ${d_0 \delta_1 /(4\sqrt{2} r^2 r_g c)}$, and the horizontal axis in units  ${({\omega}r_g/c)}$.}}
\end{figure*}

Rewrite (\ref{4-3}) in the linear approximation with respect to $\delta$ in the form: 
\begin{eqnarray}
c t \approx  c t_2 + \kappa\delta - r_g\, \ln (\delta) \quad \Rightarrow \quad 
\delta \approx \exp\left(\frac{ct_2 - ct}{r_g}\right)\cdot \left( 1+\frac{\kappa\delta}{r_g} \right) \approx 
\exp\left( \frac{ct_2 - ct}{r_g} \right) 
\label{4-4} 
\end{eqnarray} 
Here $t_2$ \& $\kappa$ are constants (accounting the constant $\kappa$ in this formula is the excess of required accuracy). 

Taking into account that ${a^2\approx\delta}$, we find that in the linear approximationthe remaining quantities of the formula (\ref{4-2-2}) can be taken in the zero approximation with respect to $\delta$: ${b\approx c_1}$, ${r_d\approx r_g}$. 
Hence the second integral (\ref{4-2-1}) (from $t_1$ to infinity) in the linear approximation with respect to small value  
$\delta$ can be written as:  
\begin{eqnarray} 
{\bf F_{{\omega}d_\perp}^{\theta t}}{}_{(II)} \approx  
\int\limits_{t_1}^{\infty} \frac{d_0\,\delta\, \left[\sin ({\omega} t) + \cos ({\omega} t)\right]}{8 r^2 r_g^2}\, dt \approx 
\int\limits_{t_1}^{\infty} \frac{d_0\, \exp\left[c(t_2-t)/r_g \right]\cdot 
\left[\sin ({\omega} t) + \cos ({\omega} t)\right]}{8 r^2 r_g^2}\, dt \quad
\label{4-5}
\end{eqnarray}
Integrating, we obtain: 
\begin{eqnarray} 
{\bf F_{{\omega}d_\perp}^{\theta t}}{}_{(II)} = \frac{d_0\, \delta_1 \left[ 
\sin ({\omega} t_1) \left(1 - {\omega}r_g/c\right) + \cos ({\omega} t_1)\left(1 + {\omega}r_g/c \right)\right] 
}{8 r^2 r_g c \left[1+({\omega}r_g/c)^2\right]} = 
\frac{d_0\, \delta_1 \sin ({\omega} t_1 +\gamma)}{4\sqrt{2} r^2 r_g c \sqrt{1+({\omega}r_g/c)^2}} \quad 
\label{4-6}
\end{eqnarray}
Here the phase $\gamma$ is given by ${\sin\gamma\equiv (1 + {\omega}r_g/c)/\sqrt{2[1+({\omega}r_g/c)^2]}}$. 

Attractor of expression (\ref{4-6}) is a function: 
\begin{eqnarray} 
Att \equiv \frac{d_0\, \delta_1}{4\sqrt{2} r^2 r_g c \sqrt{1+({\omega}r_g/c)^2}} \to 
\frac{d_0\, \delta_1}{4\sqrt{2} r^2 r_g c ({\omega}r_g/c)} 
\label{4-7}
\end{eqnarray}
Here the arrow indicates the limit at ${({\omega}r_g/c)\to\infty}$. 
The comparison of attractor with the numerical value of the second part of the integral (\ref{4-2-1}) is shown in Figure~\ref{R3}. 

The index $k$ for the radiation spectral energy density corresponding to the attractor asymptotic behavior (\ref{4-7}) is equal to minus two\footnote{Because the spectral energy density is the square of the Fourier component of the field.}.  
But we still need to include the contribution from the first part of the integral (\ref{4-2-1}), which (as will be seen below) is the main contribution to the integral. 

\begin{figure*}
\subfigure[\label{R3-1} ${r_0=4r_g}$]{\includegraphics[width=0.49\textwidth,  height=0.14\textheight]{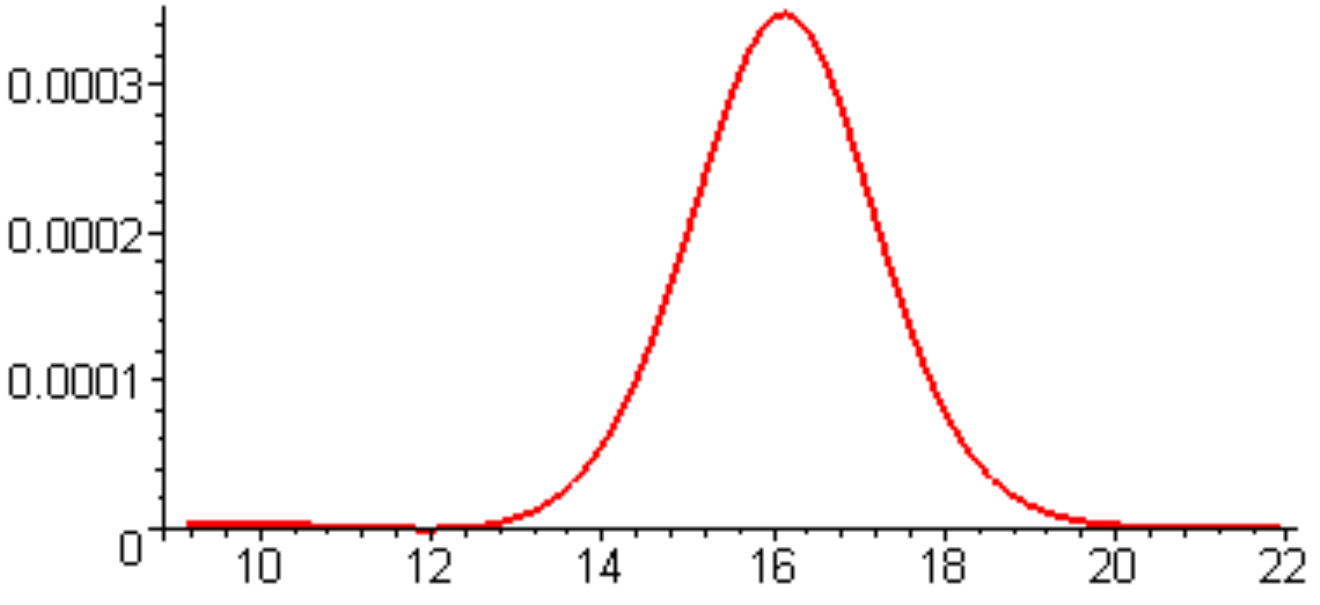}} 
\subfigure[\label{R3-2} ${r_0=10r_g}$]{\includegraphics[width=0.49\textwidth,  height=0.14\textheight]{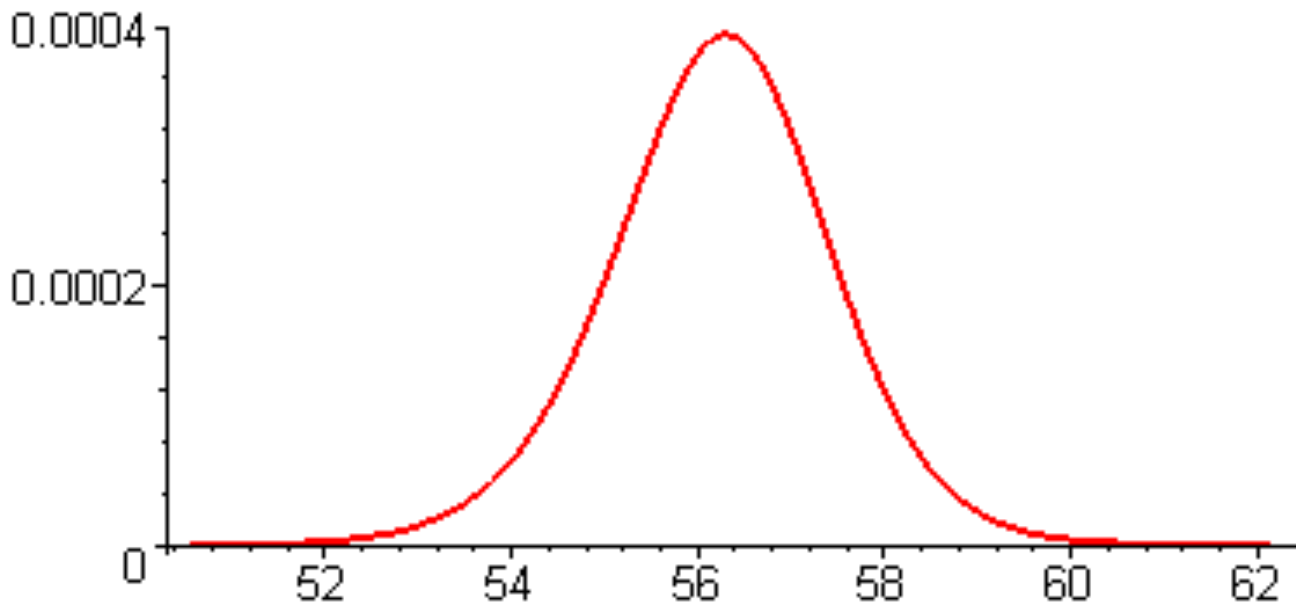}} 
\subfigure[\label{R3-2} ${r_0=20r_g}$]{\includegraphics[width=0.49\textwidth,  height=0.14\textheight]{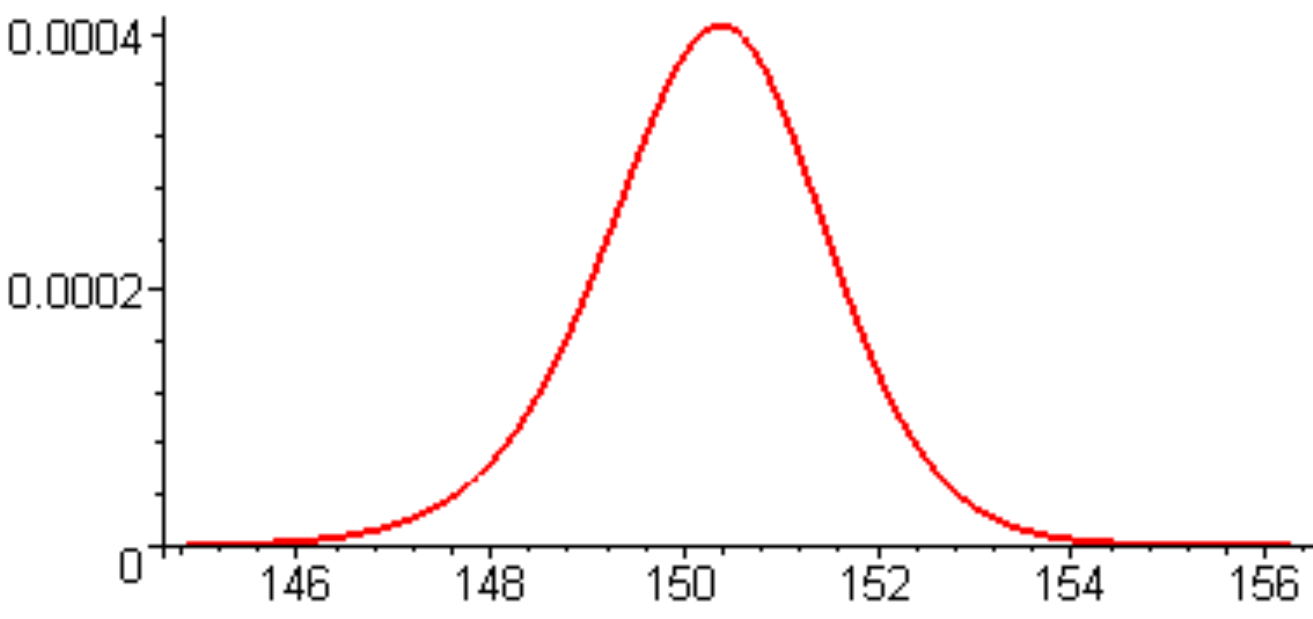}} 
\subfigure[\label{R3-4} ${r_0=100r_g}$]{\includegraphics[width=0.49\textwidth,  height=0.14\textheight]{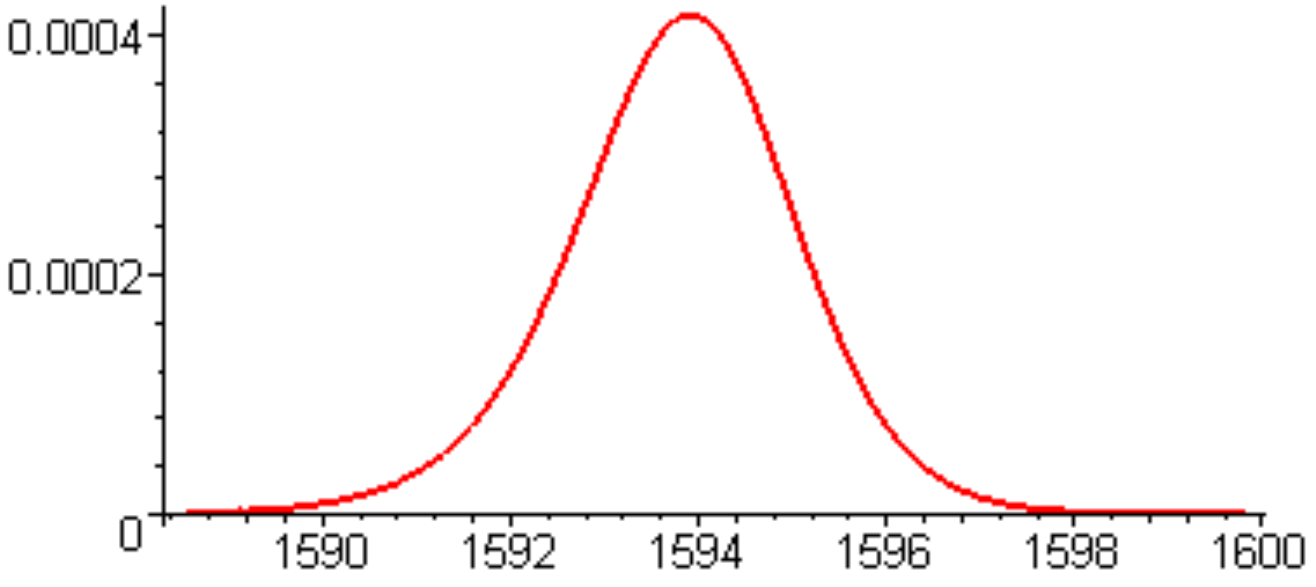}} 
\caption{{\label{R4} Dependences for the square of the radiation field ${(F^{\theta t}_{d_\perp})^2}$ (proportional to the Poynting vector) on time --- see~(\ref{4-2-2}, \ref{4-3}).  
The vertical axis is scaled in units ${d_0^2/(r r_g)^4}$ and horizontal --- scaled in units ${(ct/r_g)}$.  
At the maximum ${(F^{\theta t}_{d_\perp})^2\equiv F_m^2}$. 
}} 
\end{figure*}

Expression (\ref{4-2-2}) has a pronounced local extreme. 
This extremum can be analytically determine exactly, but there is no need to do this. 
Since time $t$ monotonically associated with the radius of the dipole $r_d$ by the law of motion~(\ref{4-3}), 
then the expression (\ref{4-2-2}) can also be represented as a function of time --- see Fig.~\ref{R4}.

Assume that the point of this extremum is located at ${t=t_m}$ and ${F^{\theta t}_{d_\perp}(t_m)\equiv F_m}$. 
We expand the function ${F^{\theta t}_{d_\perp}(t)}$ near the $t_m$ into a Taylor series in magnitude ${(t-t_m)}$:  
\begin{eqnarray} 
F^{\theta t}_{d_\perp}(t) = F_m + {C_2} (t-t_m)^2 + {C_3} (t-t_m)^3 + ... 
\label{4-9}
\end{eqnarray} 
This expansion is valid we suppose in finite time interval: ${t\in [0, t_1]}$.  
Substituting~(\ref{4-9}) in the first part of the integral~(\ref{4-2-1}), we obtain: 
\begin{eqnarray} 
{\bf F_{{\omega}d_\perp}^{\theta t}}_{(I)} = \frac{1}{2{\omega}}
\int\limits_{0}^{x_1} \left[\sin (x) + \cos (x)\right]\cdot 
\left[ F_m + \frac{{C_2} (x-x_m)^2}{{\omega}^2} + \frac{{C_3} (x-x_m)^3}{{\omega}^3} + ... \right]\, dx  
\label{4-10}
\end{eqnarray} 
Here we introduce the variables ${x\equiv {\omega}t}$, ${x_1\equiv {\omega}t_1}$ and ${x_m\equiv {\omega}t_m}$. 

Therefore, in the limit of ${({\omega} >> c/r_g)}$, we obtain: 
\begin{eqnarray} 
{\bf F_{{\omega}d_\perp}^{\theta t}}{}_{(I)} \approx \frac{F_m}{2{\omega}}
\int\limits_{0}^{x_1} \left[\sin (x) + \cos (x)\right]\, dx = \frac{F_m}{2{\omega}} \left[ 1+\sin(x_1)-\cos(x_1) \right] 
= \nonumber\\ 
= \frac{F_m}{\sqrt{2}{\omega}} \left[ \frac{1}{\sqrt{2}} + \sin\left({\omega} t_1 - \frac{\pi}{4} \right) \right]
\label{4-11}
\end{eqnarray}
It follows that for the first part of the integral (\ref{4-2-1}) ${{\bf F_{{\omega}d_\perp}^{\theta t}}{}_{(I)}}$ 
index $k$ for the spectral radiation energy density is also equal to minus two. 
Moreover with ${\delta_1 << 1}$ the expression (\ref{4-11}) is a major contribution in the integral (\ref{4-2-1}), because  
${{\bf F_{{\omega}d_\perp}^{\theta t}}{}_{(I)} >> {\bf F_{{\omega}d_\perp}^{\theta t}}{}_{(II)}}$.  

Thus for ${({\omega} >> c/r_g)}$ the statement that ${k = -2}$ holds for all integral (\ref{4-2-1}).

\section{The characteristic observation features of the spectrum}
\label{s7}

A distinctive feature of the radiation spectral energy density from dipole which falling into the black hole is not so much its asymptotic attractor as a characteristic oscillation period of the spectrum --- see Fig.~\ref{R1}.

\begin{figure*}
\includegraphics[width=0.7\textwidth, height=0.2\textheight]{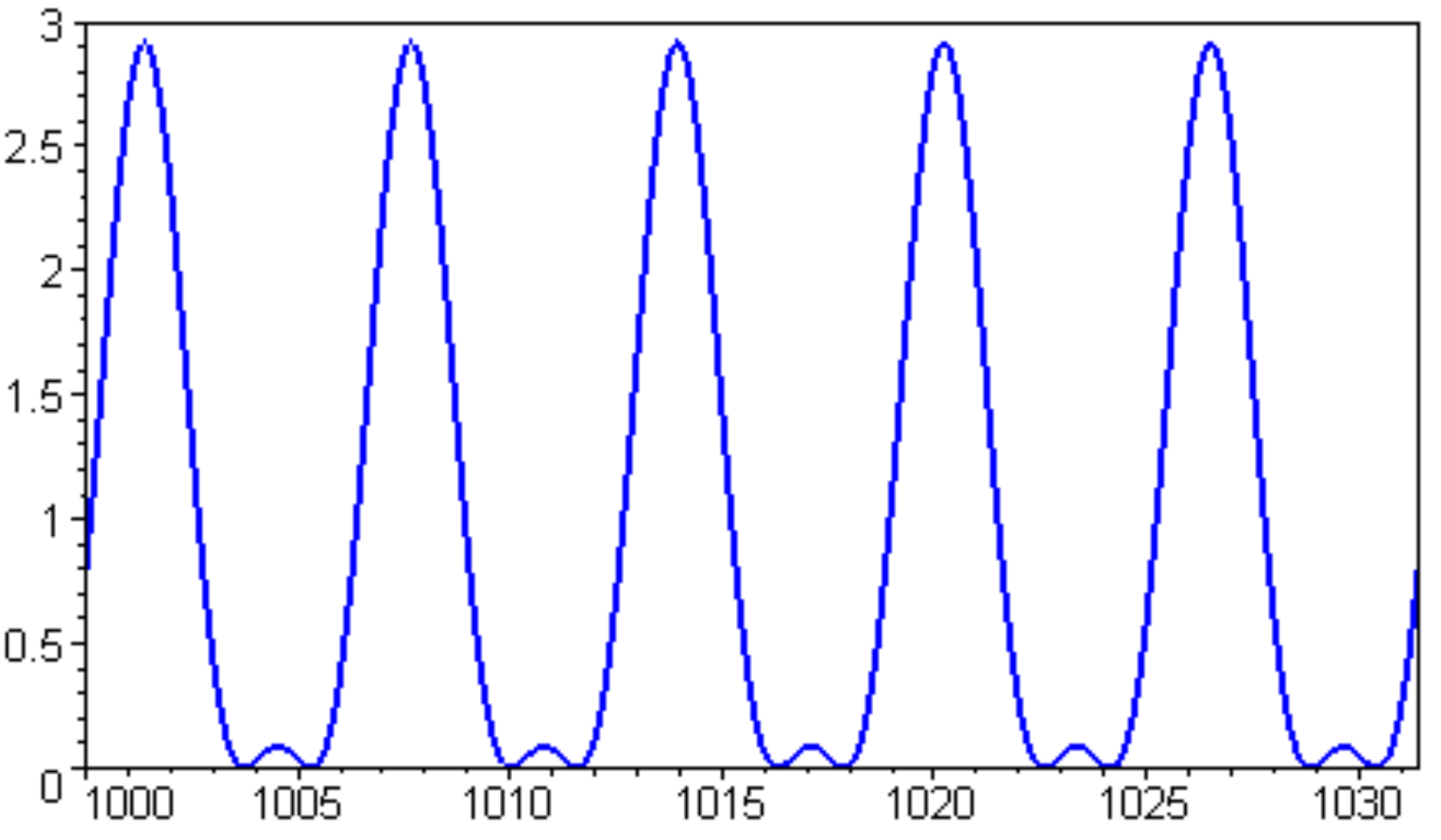}
\caption{{\label{R5}The characteristic shape of the spectrum. 
Plot of the expression in the square brackets from the formula~(\ref{4-12}). 
On the horizontal axis --- argument ${({\omega}t_1)}$.}} 
\end{figure*}

According to~\cite{Landau1988}, \S 66 the spectral energy density ${{\cal E}^d_{,{\omega}}}$ has the form\footnote{In  formula~(\ref{4-0}), we assumed the corners flux distribution of radiation energy as ${(\sim\cos^2\theta)}$.}: 
\begin{eqnarray} 
{\cal E}^d_{,{\omega}} = \frac{c r^2}{3\pi} \left({\bf F_{{\omega}d_\perp}^{\theta t}}
\sqrt{|g_{\theta\theta}|}\right)^2
\label{4-0}\end{eqnarray}
Therefore, from the expression (\ref{4-11}) we have (at ${{\omega}>>c/r_g}$): 
\begin{eqnarray} 
{\cal E}^d_{,{\omega}} = \frac{c r^4 F_m^2}{6\pi{\omega}^2} \left[ 1 + \sqrt{2}\sin\left({\omega} t_1 - \frac{\pi}{4} \right) 
-\frac{1}{2}\sin\left(2{\omega} t_1\right) \right]
\label{4-12}
\end{eqnarray}
According to this expression we obtained that the spectral energy density oscillates at the two frequencies (see Fig.~\ref{R5}): 
${(\Delta {\omega})_1 = \pi /t_1}$ and ${(\Delta {\omega})_2 = \pi /(2t_1)}$. 
Therefore, the definition of these frequencies (oscillations of the spectrum) is the main task of of this work. 

According to the formula (\ref{4-4}) time $t_1$ can be defined as: 
\begin{equation}
ct_1 \equiv ct_2 + r_g\ln(1/\delta_1)
\label{7-4}\end{equation}
And time $t_2$, according to (\ref{4-3}), defined as:  
\begin{eqnarray}
c t_2 \equiv (r_0-r_g) + \frac{(r_0+2r_g)\sqrt{r_0/r_g-1}}{2}\cdot \arccos\left( \frac{2r_g}{r_0} - 1\right)   
+ r_g\cdot\ln \left[ 4\left( 1 - \frac{r_g}{r_0}\right) \right] 
\label{7-5}
\end{eqnarray} 
At ${r_0/r_g >\sim 10}$ we have: ${c t_2>>r_g\ln(1/\delta_1)}$, in this case\footnote{But for ${r_0/r_g\le\sim 10}$ need to use the exact expressions (\ref{7-4}-\ref{7-5}).} 
\begin{eqnarray}
c t_1 \approx c t_2 \approx (r_0 - r_g) + \frac{\pi (r_0+2r_g)\sqrt{r_0/r_g - 1}}{2}   
\label{7-6}
\end{eqnarray} 
The numerical calculation and graphical representation (see Fig.~\ref{R4}) shown that the characteristic radiation of 
${\Delta t}$ (which can be estimated as the length of radiation with a power greater than the half-maximal) is virtually independent from the initial radius $r_0$. 
We have: ${{\Delta t}\approx 3r_g/c}$. 
Then, knowing the time of the observations ${\Delta t}$, we can calculate the value of ${r_g\approx c{\Delta t}/3}$ and the black hole mass. 

Expression (\ref{7-6}) can be reduced to a cubic equation for ${r_0}$. 
The value ${t_1}$ also determined from observations: ${t_1 = \pi /(\Delta {\omega})_1}$. 
Therefore calculating the roots of the cubic equation it will be possible to estimate the initial distance $r_0$ of dipole to the black hole before falling. 
In the more general case (for example, when ${r_0/r_g\le\sim 10}$) the value of $r_0$ can also be determined numerically. 

In addition, the amplitude of the radiation power (which is also almost does not depend on initial radius $r_0$) will be available the ratio of the dipole moment to the distance from black hole: ${d_0/r}$.

\section{Observational constraints} 
\label{s3} 

In order to be able to register the falling radiation from the dipole it is necessary that the spectral energy density 
${\cal E}^d_{,{\omega}}$, which divided by time ${{\Delta t}\approx 3r_g/c}$ (during which the dipole radiates --- see 
Fig.~\ref{R4}), will be greater than the maximum sensitivity of the instrument $J^{dev}$, multiplied by the total area of the scattering of radiation ${4\pi r^2}$ (here $r$ --- the distance from the black hole to the observer, ${r>>r_0}$): 
\begin{equation} 
{\cal E}^d_{,{\omega}} /{\Delta t} > 4\pi r^2 J^{dev} 
\quad\makebox{or}\quad 
J\equiv {\cal E}^d_{,{\omega}} /(4\pi r^2 {\Delta t}) > J^{dev} 
\label{3-0}\end{equation} 
Assume for definiteness that that the maximum sensitivity of the instrument is  
${J^{dev}=10^{-3} [Jy]}$, and ${1[mJy] = 10^{-26} [erg/(sm^2\, sec\, Hz)].}$ 

In according to the Fig.~\ref{R4} we obtain approximately ${|F_m|\approx 0.02 d_0/(rr_g)^2}$. 
Therefore, when ${{\omega}>>c/r_g}$ and in accordance with the expressions (\ref{4-12}) and (\ref{3-0}), we have:  
\begin{eqnarray} 
{\cal E}^d_{,{\omega}} \approx \frac{10^{-4}}{{\omega}^2}\cdot\frac{2 c d_0^2}{3\pi r_g^4}  
\quad\makebox{or}\quad 
J\approx \frac{10^{-4}}{{\omega}^2}\cdot\frac{c d_0^2}{6\pi^2 r^2 {\Delta t} r_g^4} \, . 
\label{7-1}
\end{eqnarray}
According to~(\ref{3-0}), we obtain restrictions on the frequency that you can still register with the receiver sensitivity $J^{dev}$: 
\begin{equation}
{\omega} < \frac{d_0}{r r_g^2}\cdot\sqrt{\frac{10^{-4} c}{6\pi^2 {\Delta t} J^{dev}}}
 \approx \frac{d_0 c}{100\pi r r_g^2 \sqrt{18 r_g J^{dev}}}
\label{7-3}\end{equation}
Let us estimate the dipole value $d_0$, which is necessary for detection of radiation at frequency ${{\omega}>>c/r_g}$ when it falls into a black hole from radius ${r_0=10r_g}$. 
We assume that the magnetic dipole radiates with the same spectrum as the electric. 
Physically, the magnetic dipole in the first approximation can be a pulsar or a planet with its own magnetic field. 
If this is a pulsar, the magnetic field at the surface is ${H_{puls}\sim 10^{12} [G]}$, and the size of the pulsar is  
${R_{puls}\approx 2\cdot 10^6 [sm]}$. 
Hence ${d_0^{puls}\approx H_{puls}\cdot R_{puls}^3\approx 10^{31} [SGS]}$. 

Similarly, for a planet like the Earth we receives: ${R_{plan}\approx 6.4\cdot 10^8 [sm]}$, 
${H_{plan}\approx 0.5 [G]}$ and ${d_0^{plan}\approx 10^{26} [SGS]}$.

From the Fig.~\ref{R1} and the formula (\ref{7-1}) we see that maximum spectral energy density is about 
${{\cal E}^d_{,{\omega}(max)}\approx 3\cdot 10^{-4} d_0^2/(c r_g^2)}$, and the corresponding value  
${J_{max}\approx 6\cdot 10^{-6} d_0^2/(r^2 r_g^3)}$ --- at frequencies of about ${{\omega}_{max}\approx 0.2c/r_g}$. 

For example, for the pulsar, which falling into a black hole (which is located at a distance 
${r\sim 8 [kPc]\approx 2.4\cdot 10^{22} [sm]}$ from the Sun --- in the center of the Milky Way) and have the mass  
${M_{bh}\sim 10^6 M_\odot}$, we obtain at the maximum radiation: ${J^{puls}_{max}\approx 100 [Jy]}$ --- at a frequency  
${{\omega}_{max}^{puls}\approx 0.1 [Hz]}$. 
Then by formula (\ref{7-3}) we obtain for the pulsar: ${{\omega}^{puls}<\sim 30\, [Hz]}$. 
Ie at frequencies above the ${\sim}$30Hz the radiation flux from the pulsar which falling in the central (in our galaxy) black hole is less than one mJy. 

Respectively, for a magnetized planet or asteroid with a magnetic dipole $d_0^ {plan}$ (like Earth), falling into a black hole, which is located at a distance of ${r\sim 100[Pc]}$ from the sun and has a mass of ${M_{bh}\sim 10^3 M_\odot}$, we obtain at the maximum radiation: ${J^{plan}_{max}\approx 5\cdot 10^{4} [Jy]}$ --- at a frequency   
${{\omega}_{max}^{puls}\approx 60 [Hz]}$. 
Then by formula (\ref{7-3}) we obtain for the planet: ${{\omega}^{plan}<\sim 900 [kHz]}$. 

For this reason, the study of magnetized matter which falling into black holes with intermediate mass seems most
interesting.
The process of falling of large bodies into black holes with a mass of less than ${\sim 10^3 M_\odot}$ must be considered carefully, because in this case we neglect the important limitation of our model --- dipole size should be much smaller than
black hole (otherwise the dipole can be destroyed by tidal forces).

\section{Discussion and conclusions}
\label{s8}

In the case of detection such observation spectra will be a real allow new and independent way to determine the main characteristics of the black hole --- its mass. 
In addition, is also likely to be possible to determine by circumstantial evidence, some properties of magnetized matter which accreting onto the black hole.   

The main problem which arises when, would be too weak (for measuring) energy flow in the radio frequencies for observation of a such process (dipole fall into a black hole). 

In addition, apparently the fall of the strongly magnetized compact bodies to the black holes are very rare event which is also a serious obstacle for the observations.  

But it is also likely that the scanning of entire range in radioastronomy (for searching the similar spectra for such as we have) will help to discover new black holes in our Galaxy and will help the study of their properties.

\section*{Aknowledgments}
\label{s9}

We are particularly grateful to N.S. Kardashev, K.A. Bronnikov and and all workshops participants for many useful
discussions on the subject and for valuable comments.

This work was supported by RFBR, project codes: 12-02-00276-a,
11-02-00244-a, 11-02-12168-ofi-m-2011, Scientific
School-2915.2012.2 "Formation of large-scale structure of the
Universe and cosmological processes" Programme "Nonstationary
Phenomena in the objects of the universe 2012"\, and the Federal
Target Program "Scientific and pedagogical Staff of Innovative
Russia 2009-2013" \, 16.740.11.0460.

\end{document}